\documentclass{article}

\usepackage[dblblindworkshop, final]{neurips_2025}

\usepackage[utf8]{inputenc} 
\usepackage[T1]{fontenc}    
\usepackage{hyperref}       
\usepackage{url}            
\usepackage{booktabs}       
\usepackage{amsfonts}       
\usepackage{nicefrac}       
\usepackage{microtype}      
\usepackage{xcolor}         
\usepackage{colortbl}
\usepackage{graphicx}

\newcommand{\squishlist}{
   \begin{list}{$\bullet$}
    { \setlength{\itemsep}{0pt}      \setlength{\parsep}{3pt}
      \setlength{\topsep}{3pt}       \setlength{\partopsep}{0pt}
      \setlength{\leftmargin}{1.0em} \setlength{\labelwidth}{1em}
      \setlength{\labelsep}{0.5em} } }
\newcommand{\squishend}{
    \end{list}  }

\title{Rethinking Kernel Program Repair: Benchmarking and Enhancing LLMs with RGym}

\author{%
  Kareem Shehada \\
  University of California, Riverside\\
  \texttt{kareem.shehada@email.ucr.edu}\\
  \And 
  Yifan Wu \\
  University of California, Riverside\\
  \texttt{yifan.wu1@email.ucr.edu}
  \And 
  Wyatt D. Feng \\
  University of California, Riverside\\
  \texttt{wyatt.feng@email.ucr.edu} \\
  \And 
  Adithya Iyer \\
  University of California, Riverside\\
  \texttt{aiyer026@ucr.edu}
  \And 
  Gryphon Kumfert \\
  University of California, Riverside\\
  \texttt{gkumf001@ucr.edu}\\
  \And 
  Yangruibo Ding \\
  Columbia University\\
  \texttt{yrbding@cs.columbia.edu}\\
  \And 
  Zhiyun Qian \\
  University of California, Riverside\\
  \texttt{zhiyunq@cs.ucr.edu}
}

\begin{document}

\workshoptitle{Evaluating the Evolving LLM Lifecycle: Benchmarks, Emergent Abilities, and Scaling}

\maketitle

\begin{abstract}
Large Language Models (LLMs) have revolutionized automated program repair (APR) but current benchmarks like SWE-Bench predominantly focus on userspace applications and overlook the complexities of kernel-space debugging and repair. The Linux kernel poses unique challenges due to its monolithic structure, concurrency, and low-level hardware interactions. Prior efforts such as KGym and CrashFixer have highlighted the difficulty of APR in this domain, reporting low success rates or relying on costly and complex pipelines and pricey cloud infrastructure. In this work, we introduce RGym, a lightweight, platform-agnostic  APR evaluation framework for the Linux kernel designed to operate on \emph{local commodity hardware}. Built on RGym, we propose a simple yet effective APR pipeline leveraging specialized localization techniques (e.g., call stacks and blamed commits) to overcome the unrealistic usage of oracles in KGym. We test on a filtered and verified dataset of 143 bugs. Our method achieves up to a 43.36\% pass rate with GPT-5 Thinking while maintaining a cost of under \$0.20 per bug. We further conduct an ablation study to analyze contributions from our proposed localization strategy, prompt structure, and model choice, and demonstrate that feedback-based retries can significantly enhance success rates.
\end{abstract}
\vspace{-0.1in}
\section{Introduction}
\vspace{-0.05in}

Large language models (LLMs) are rapidly reshaping software development workflows, from code generation, simple debugging, to fully automated program repair (APR)~\cite{yang2024sweagent, wang2025openhands,interfix,xu2025aligningllm, xiang2024farpracticalfunctionlevelprogram, repairagent,xia2024conversation, nong2025appatchautomatedadaptiveprompting,kulsum2024reasoning}. While existing benchmarks, such as SWE-Bench~\cite{jimenez2024swebench}, have driven steady progress on developing prototypes for LLM-based APR, their settings and samples focus on the user-space applications and underrepresent challenges common in more complicated and security-critical operating system kernel space: the kernel could potentially concentrate the hardest failure modes of systems programming with its massive scale, deep dependency, and pervasive concurrency and low-level interactions with hardware.
These characteristics make the kernel an ideal stress test for evaluating LLM-based APR, from localization to patch generation, validation, and cost/latency consideration.

Syzkaller~\cite{syzkaller}, a coverage-guided kernel fuzzer, together with Syzbot \cite{syzbot}, an automated online crash reporting system developed by Google, provides a valuable ecosystem that makes kernel-bug collection possible (more background in Appendix \ref{sec:background} ), and based on which, kGym~\cite{mathai2024kGym} introduced a platform and dataset to benchmark LLMs on Linux kernel crash resolution. Unfortunately, however, kGym's kernel gym has a hard dependency on GCP (Google Cloud Platform) and cannot be run elsewhere, restricting budget and flexibility. Furthermore, kGym uses whatever dependencies and compiler version are provided by the distribution package manager. This can easily cause build failures and can subtly change the behavior of the produced binary. To address these limitations, we introduce RGym, a lightweight, platform-agnostic solution built for local commodity hardware. RGym solves the compiler and dependency problem by smartly switching build dependencies using docker images depending on the kernel version or compiler string provided in the kernel configuration.


Besides the gym framework, \cite{mathai2024kGym} also provided a basic APR solution. With the state-of-the-art LLMs, such as GPT-4, kGym's APR approach achieved a success rate of only 0.72\% and 5.38\% in unassisted and oracle-assisted modes, respectively. 
Recently, CrashFixer~\cite{mathai2025crashfixercrashresolutionagent} followed up with a more complex design of APR, using a debug tree to generate hypotheses of root causes and iteratively refining them into patches, which led to an oracle-assisted pass rate of 65.6\% at a high cost of \$21.62 per bug. 

Contrary to the difficulties suggested by prior work, we find that simpler APR designs can achieve results comparable to CrashFixer while relying on more realistic assumptions and incurring significantly lower costs. Our main findings are as follows. First, both kGym and CrashFixer assume access to oracles for identifying the relevant files to patch, which is unrealistic in practice; in contrast, we demonstrate that practical localization strategies can achieve strong results, such as providing a bug inducing commit \cite{wen2020boostbic} that hints the root cause, which is obtainable using recent advances in bug bisection solutions targeting Syzbot bugs \cite{zheng2024symbisect}. Second, with relatively straightforward designs combining realistic localization with other known techniques, we achieve pass rates of 37.76\% and 43.36\% using GPT-4o and GPT-5 (Thinking model), respectively, at costs of only less than \$0.2 per bug. Third, we conduct a detailed ablation study that isolates the contributions of different components in our pipeline, including the localization strategy, prompt structure, and choice of LLM models. 
Lastly, we find that different design choices/configurations of the solution can often complement each other, highlighting the benefits of diversification.

\squishlist

\item 
We introduce a patch testing system called RGym. RGym automatically handles build and test dependencies to streamline testing and reduce the domain knowledge required to adequately test APR tools. RGym is designed to be easy to set up locally.

\item 
We organize a dataset of 143 kernel bugs from Syzbot into an easily consumable format and verified the reproducibility of the bug on the patch parent. These kernel bugs have developer-curated bug-inducing commits, facilitating the ground truth for localization.

\item 
We develop a simple yet more effective APR than kGym and propose a different method of localization using bug-inducing commits and call stack. The results achieve pass rates of 37.76\% to 43.36\% using different LLM models --- the combined pass rates reach 68.53\%. We conducted an ablation study to measure the impact and cost of different components, such as parts of the prompt and the LLM model used.

\squishend
\section{Methodology}
\label{sec:methodology}

\begin{figure}[!t]
  \centering
  \includegraphics[width=.8\textwidth]{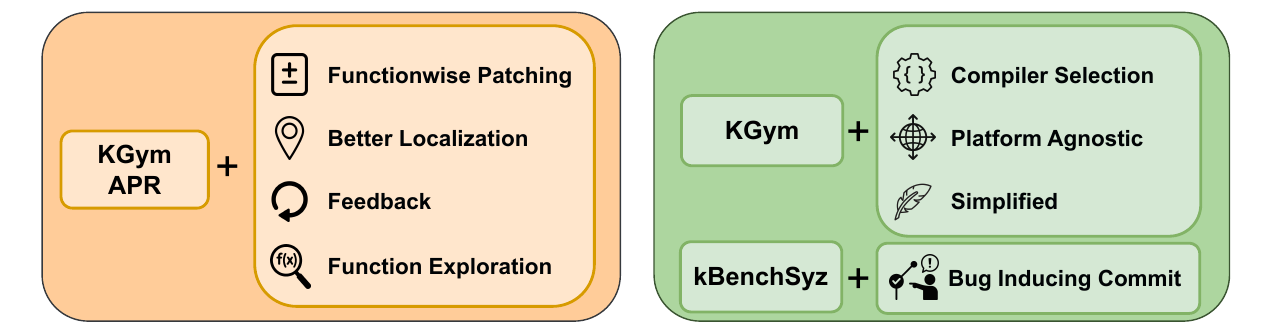}
  \caption{RGym's improvements and additions over kGym's APR, gym, and dataset.}
  \label{fig:improvements}
\end{figure}

Our system, as shown in Figure  ~\ref{fig:improvements}, is composed of two main components: RGym, a testing framework, and an APR tool. The APR generates a patch via the Simple Agent or Function Exploration Agent and tests it with RGym. On failure, a feedback module can be leveraged to summarize the issue and retry. We evaluate on a dataset of 143 verified bugs.

\textbf{Dataset:}
From 6,088 Syzbot bugs, we retain those with fix commits, reproducers, crash reports, and kernel configs, filtering to KASAN bugs \cite{kasan}, which represent the most severe types of bugs (memory corruption)~\cite{zou2022syzscope,KOOBE}. 

Using RGym, we also verify reproducibility at the parent of the fix commit. 
This leads to 143 reproducible KASAN bugs, including out-of-bounds memory access, use-after-free, and null-pointer-dereference bugs.

\textbf{RGym:}
RGym overall compiles patched kernels, runs PoCs, and reports results. Unlike kGym’s cloud-based setup, RGym runs locally using docker to bundle job dependencies and QEMU for VMs. It exposes a web API and Python library for managing jobs, results, and logs.

\emph{Build job:}
It compiles the patched kernel from inputs (patch, commit, source, config, compiler, cores, timeout, metadata). The prebuilt Debian images mitigate dependency and compiler version issues encountered when building the kernel. Outputs are a kernel image or the type of failure.

\emph{Reproducer job:}
It boots a VM with the patched kernel and Debian rootfs to run syz/C reproducers. Inputs include kernel image, reproducer, timeout, cores, and metadata. Returns success on timeout, or the type of failure.

\textbf{APR tool:}
Our APR is composed of two agents: The Simple Agent that provides example patches (via in-context learning) for OOB, UAF, and NPD bugs. The Function Exploration Agent can perform on-demand code viewing to develop its own view of the bug root cause, and therefore the corresponding patch strategy may differ. 
Both agents use the BIC-based localization (together with callstack). Both use GPT-4o as the baseline for cost efficiency.

\emph{Function-wise patching:} 
The LLM lists candidate functions, receives their definitions, and returns their patched definitions. All changes are encompassed into a single diff, ensuring applicability without concerns about diff syntax.

\emph{Realistic localization:} 
Unlike assuming the knowledge of which files to patch (oracles), our localization depends on BICs, which have been demonstrated as achievable. Specifically, SymBisect \cite{SymBisect_opensource} provided an automated approach to identify BICs in Syzbot bugs, achieving 75\% accuracy.

\emph{Retries and error summary:}
On failure (e.g., build error, sanitizer trigger) the APR asks LLM to summarize the issue, then restarts the agent with the summary appended.

\emph{Function Exploration:}
This design allows LLMs to freely request additional function definitions, enabling them to build a localized view of the potential root cause instead of being limited to specialized prompts (and bug types).

\section{Evaluation}
\label{sec:evaluation}

In this section, we evaluate our approach to automated program repair (APR). All patches, once built, are tested with 26 VMs running the reproducer(s) (either 13 Syz, 13 C; or 26 Syz) in a loop for 10 minutes. We do this evaluation with respect to three key research questions (RQs):
\squishlist
\item \textbf{RQ1}: How do our included APR components improve patch pass rates?

\item \textbf{RQ2}: What are the costs of each APR configuration and how do the costs compare to their effectiveness?

\item \textbf{RQ3}: How do SOTA LLM models perform using our APR and are they cost-effective? 

\squishend

\subsection{RQ1: Effect of APR components on repair success}

\noindent\textbf{kGym and function-wise patching.} We summarize the key results in Table~\ref{tab:overall-results}. We first revisit kGym's reported pass rates. kGym evaluates each candidate patch by rerunning the reproducer in a single VM continuously for 10 minutes. However, in practice, many bugs are stateful and many reproducers are non-deterministic: In Table \ref{tab:reproducer-job} we observe that roughly one-third of bugs have non-deterministic reproducers, leading to unreliable triggering. Using the oracle (knowing which file should be patched), kGym's reported 5.38\% pass rate shrinks to 1.4\% because of this.

For kGym, bad patches (those that fail to apply) account for most failed attempts and build errors, as LLMs often struggle to generate precise diffs (e.g., correct line numbers). 

When we introduce function-wise patching to kGym's APR, we see a significant mitigation of the problem. Bad patches are reduced by 76\%, in turn increasing overall success from 2.8\% to 10.49\%, underlining the necessity of dedicated patching components to complement raw LLM outputs.

\noindent\textbf{Localization, function exploration, and feedback.} We then transition to our Simple Agent APR using bug-type specific instructions and call stack localization (without feeding the BIC), neither of which requires oracle guidance as they're sourced from the sanitizer report. This configuration achieves 17.48\% pass rate, a 6.99\% improvement over kGym's oracle-guided solution with function-wise patching. Adding the BIC to complement call stack localization pushes the pass rate to 21.67\%, a 4.19\% improvement. While the BIC is generally not available for unpatched bugs, tools like SymBisect \cite{zheng2024symbisect} can obtain the BIC with 75\% accuracy. Our non-bug type-specific agent, Function Exploration Agent, achieves a 15.38\% pass rate, but provides a decent complement to Simple Agent. Of the 22 bugs patched, 12 are uniquely solved by our Function Exploration Agent, giving a combined pass rate of 30\%.
Our Simple Agent with feedback enabled and up to 3 retries achieves a 37.76\% pass rate. We see that 34 bugs (23.77\%) are solved in the first attempt, 8 (5.59\%) in the second attempt, and 12 (8.39\%) in the third attempt. These results show there is value in retrying even beyond three attempts; however, the benefit is diminishing.

\begin{table}[]
\caption{Overall Results}
\label{tab:overall-results}
\centering
\begin{tabular}{lllll}
\toprule
Setup                    & LLM             & Pass Rate & Bad Patch & Avg \$/Bug \\
\midrule
kGym-oracle              & GPT-4-turbo     & 1.4\%     & 59.43\%   & 0.21       \\
kGym-oracle              & GPT-4o          & 2.8\%     & 51.88\%   & 0.05       \\
kGym-oracle+functionwise & GPT-4o          & 10.49\%   & 12.14\%   & 0.06       \\
SimpleAgent-nobic        & GPT-4o          & 17.48\%   & 1.39\%    & 0.05       \\
SimpleAgent              & GPT-4o          & 21.67\%   & 4.89\%    & 0.08       \\
SimpleAgent+Feedback     & GPT-4o          & 37.76\%   & 4.89\%    & 0.17       \\
ExplorationAgent     & GPT-4o          & 15.38\%   & 5.59\%    & 0.12       \\
SimpleAgent              & Claude Opus 4.1 & 32.16\%   & 5.59\%    & 0.73       \\
SimpleAgent              & GPT-5 Thinking  & 43.36\%   & 4.19\%    & 0.18    \\
\bottomrule
\vspace{-.1in}
\end{tabular}
\end{table}

\subsection{RQ2: Costs of each APR configuration compared to effectiveness}
 
As shown in Table~\ref{tab:overall-results}, kGym with GPT-4o costs only \$0.05 per bug in oracle mode.
 
Our subsequent improvements only mildly increase the costs.

Our Simple Agent with BIC costs \$0.08 per bug. 
Our Function-Exploration Agent costs \$0.12 per bug, which is somewhat expensive for its lower pass rate. However, it is still useful given its complementary nature.

The average cost per bug of Simple Agent with feedback (3 tries) is \$0.17, 2.13x the cost of running Simple Agent once, while achieving 1.74x the pass rate.

\subsection{RQ3: SOTA LLM models and their effectiveness}

As shown in Table~\ref{tab:overall-results}, our Simple Agent using Claude Opus 4.1 reaches a 32.16\% pass rate, while costing \$0.73 per bug. 

Our Simple Agent using GPT-5 Thinking achieves an impressive 43.36\% pass rate at \$0.18 per bug. This is a 5.6\% improvement over Simple Agent using feedback/retry, while costing only 1 cent more per bug. GPT-5 Thinking clearly outperforms Claude Opus 4.1 in this test, costing 4.05x less while performing 11.2\% better.
CrashFixer achieves 65.6\% pass rate at a cost of \$21.62 per bug using Gemini 2.5 Pro on kGym's kBenchSyz dataset, which is similar enough to our dataset to make some analysis. CrashFixer is 120.11x more expensive than SimpleAgent using GPT-5 Thinking, while performing only 22.24\% better despite using oracle-guided localization. If we consider the combined pass rates (union of solved bugs) of our configurations, we see a 68.53\% pass rate at an average cost of \$1.33 per bug. This leaves the question as to whether CrashFixer's complex and expensive strategy is truly necessary, but we do not perform further evaluation with CrashFixer as it is currently closed source.
\section{Conclusion}

This work introduces RGym, a lightweight, platform-agnostic evaluation framework for LLM-based automated program repair (APR) in the Linux kernel space. 
Alongside RGym, we present an effective suite of APR strategies grounded in practical localization techniques -- notably using bug-inducing commits (BICs), call stacks, and function-wise patching -- that do not rely on unrealistic oracle assumptions. Our evaluation showed that our solution can significantly improve the pass rates of generated patches, with a fairly modest cost.

\bibliographystyle{plain}
\bibliography{reference}

@String { AUG          = {August} }

@String { COMPUTER     = {IEEE Computer Magazine} }

@String { MAY          = {May} }

@Misc{syzbot,
  Author                   = {{Google}},
  Title                    = {{Google syzbot}},
  HowPublished             = {\url{https://syzkaller.appspot.com/upstream/}},
}

@Misc{syzkaller,
  Author                   = {{Google}},
  Title                    = {{Google syzkaller}},
  HowPublished             = {\url{https://github.com/google/syzkaller}},
}

@inproceedings{KOOBE,
    author = {Weiteng Chen and Xiaochen Zou and Guoren Li and Zhiyun Qian},
    title = {KOOBE: Towards Facilitating Exploit Generation of Kernel Out-Of-Bounds Write Vulnerabilities},
    series = {USENIX Security},
    year = {2020}
}

@Misc{SymBisect_opensource,
  Title                    = {{SymBisect Source Code}},
  HowPublished             = {\url{https://github.com/zhangzhenghsy/SymBisect}},
}

@inproceedings{zou2022syzscope,
  title={$\{$SyzScope$\}$: Revealing $\{$High-Risk$\}$ Security Impacts of $\{$Fuzzer-Exposed$\}$ Bugs in Linux kernel},
  author={Zou, Xiaochen and Li, Guoren and Chen, Weiteng and Zhang, Hang and Qian, Zhiyun},
  booktitle={31st USENIX Security Symposium (USENIX Security 22)},
  pages={3201--3217},
  year={2022}
}

@inproceedings{jimenez2024swebench,
    title={{SWE}-bench: Can Language Models Resolve Real-world Github Issues?},
    author={Carlos E Jimenez and John Yang and Alexander Wettig and Shunyu Yao and Kexin Pei and Ofir Press and Karthik R Narasimhan},
    booktitle={The Twelfth International Conference on Learning Representations},
    year={2024},
    url={https://openreview.net/forum?id=VTF8yNQM66},
}

@inproceedings{
yang2024sweagent,
title={{SWE}-agent: Agent-Computer Interfaces Enable Automated Software Engineering},
author={John Yang and Carlos E Jimenez and Alexander Wettig and Kilian Lieret and Shunyu Yao and Karthik R Narasimhan and Ofir Press},
booktitle={The Thirty-eighth Annual Conference on Neural Information Processing Systems},
year={2024},
url={https://openreview.net/forum?id=mXpq6ut8J3}
}

@inproceedings{
wang2025openhands,
title={OpenHands: An Open Platform for {AI} Software Developers as Generalist Agents},
author={Xingyao Wang and Boxuan Li and Yufan Song and Frank F. Xu and Xiangru Tang and Mingchen Zhuge and Jiayi Pan and Yueqi Song and Bowen Li and Jaskirat Singh and Hoang H. Tran and Fuqiang Li and Ren Ma and Mingzhang Zheng and Bill Qian and Yanjun Shao and Niklas Muennighoff and Yizhe Zhang and Binyuan Hui and Junyang Lin and Robert Brennan and Hao Peng and Heng Ji and Graham Neubig},
booktitle={The Thirteenth International Conference on Learning Representations},
year={2025},
url={https://openreview.net/forum?id=OJd3ayDDoF}
}

@article{mathai2024kgym,
  publtype={informal},
  author={Alex Mathai and Chenxi Huang and Petros Maniatis and Aleksandr Nogikh and Franjo Ivancic and Junfeng Yang and Baishakhi Ray},
  title={KGym: A Platform and Dataset to Benchmark Large Language Models on Linux Kernel Crash Resolution},
  year={2024},
  cdate={1704067200000},
  journal={CoRR},
  volume={abs/2407.02680},
  url={https://doi.org/10.48550/arXiv.2407.02680}
}

@misc{mathai2025crashfixercrashresolutionagent,
      title={CrashFixer: A crash resolution agent for the Linux kernel}, 
      author={Alex Mathai and Chenxi Huang and Suwei Ma and Jihwan Kim and Hailie Mitchell and Aleksandr Nogikh and Petros Maniatis and Franjo Ivančić and Junfeng Yang and Baishakhi Ray},
      year={2025},
      eprint={2504.20412},
      archivePrefix={arXiv},
      primaryClass={cs.SE},
      url={https://arxiv.org/abs/2504.20412}, 
}

@inproceedings {zheng2024symbisect,
author = {Zheng Zhang and Yu Hao and Weiteng Chen and Xiaochen Zou and Xingyu Li and Haonan Li and Yizhuo Zhai and Billy Lau},
title = {{SymBisect}: Accurate Bisection for {Fuzzer-Exposed} Vulnerabilities},
booktitle = {33rd USENIX Security Symposium (USENIX Security 24)},
year = {2024},
isbn = {978-1-939133-44-1},
address = {Philadelphia, PA},
pages = {2493--2510},
url = {https://www.usenix.org/conference/usenixsecurity24/presentation/zhang-zheng},
publisher = {USENIX Association},
month = aug
}

@misc{gcp-compute-pricing,
  title        = {Compute Engine: All Pricing},
  key          = {GCP},
  howpublished = {\url{https://cloud.google.com/compute/all-pricing}},
  note         = {Accessed: 2025-09-03},
  year         = {2025},
  organization = {Google LLC},
  url          = {https://cloud.google.com/compute/all-pricing}
}

@inproceedings{interfix,
author = {Jin, Matthew and Shahriar, Syed and Tufano, Michele and Shi, Xin and Lu, Shuai and Sundaresan, Neel and Svyatkovskiy, Alexey},
title = {InferFix: End-to-End Program Repair with LLMs},
year = {2023},
isbn = {9798400703270},
publisher = {Association for Computing Machinery},
address = {New York, NY, USA},
url = {https://doi.org/10.1145/3611643.3613892},
doi = {10.1145/3611643.3613892},
booktitle = {Proceedings of the 31st ACM Joint European Software Engineering Conference and Symposium on the Foundations of Software Engineering},
pages = {1646–1656},
numpages = {11},
location = {San Francisco, CA, USA},
series = {ESEC/FSE 2023}
}

@INPROCEEDINGS {xu2025aligningllm,
author = { Xu, Junjielong and Fu, Ying and Tan, Shin Hwei and He, Pinjia },
booktitle = { 2025 IEEE/ACM 47th International Conference on Software Engineering (ICSE) },
title = {{ Aligning the Objective of LLM-Based Program Repair }},
year = {2025},
volume = {},
ISSN = {},
pages = {2548-2560},
doi = {10.1109/ICSE55347.2025.00169},
url = {https://doi.ieeecomputersociety.org/10.1109/ICSE55347.2025.00169},
publisher = {IEEE Computer Society},
address = {Los Alamitos, CA, USA},
month =May}

@misc{xiang2024farpracticalfunctionlevelprogram,
      title={How Far Can We Go with Practical Function-Level Program Repair?}, 
      author={Jiahong Xiang and Xiaoyang Xu and Fanchu Kong and Mingyuan Wu and Zizheng Zhang and Haotian Zhang and Yuqun Zhang},
      year={2024},
      eprint={2404.12833},
      archivePrefix={arXiv},
      primaryClass={cs.SE},
      url={https://arxiv.org/abs/2404.12833}, 
}

@INPROCEEDINGS {repairagent,
author = { Bouzenia, Islem and Devanbu, Premkumar and Pradel, Michael },
booktitle = { 2025 IEEE/ACM 47th International Conference on Software Engineering (ICSE) },
title = {{ RepairAgent: An Autonomous, LLM-Based Agent for Program Repair }},
year = {2025},
volume = {},
ISSN = {},
pages = {2188-2200},
doi = {10.1109/ICSE55347.2025.00157},
url = {https://doi.ieeecomputersociety.org/10.1109/ICSE55347.2025.00157},
publisher = {IEEE Computer Society},
address = {Los Alamitos, CA, USA},
month =May}

@inproceedings{wen2020boostbic,
author = {Wen, Ming and Liu, Yepang and Cheung, Shing-Chi},
title = {Boosting automated program repair with bug-inducing commits},
year = {2020},
isbn = {9781450371261},
publisher = {Association for Computing Machinery},
address = {New York, NY, USA},
url = {https://doi.org/10.1145/3377816.3381743},
doi = {10.1145/3377816.3381743},
booktitle = {Proceedings of the ACM/IEEE 42nd International Conference on Software Engineering: New Ideas and Emerging Results},
pages = {77–80},
numpages = {4},
location = {Seoul, South Korea},
series = {ICSE-NIER '20}
}

@inproceedings{xia2024conversation,
author = {Xia, Chunqiu Steven and Zhang, Lingming},
title = {Automated Program Repair via Conversation: Fixing 162 out of 337 Bugs for \$0.42 Each using ChatGPT},
year = {2024},
isbn = {9798400706127},
publisher = {Association for Computing Machinery},
address = {New York, NY, USA},
url = {https://doi.org/10.1145/3650212.3680323},
doi = {10.1145/3650212.3680323},
booktitle = {Proceedings of the 33rd ACM SIGSOFT International Symposium on Software Testing and Analysis},
pages = {819–831},
numpages = {13},
location = {Vienna, Austria},
series = {ISSTA 2024}
}

@inproceedings{kulsum2024reasoning,
author = {Kulsum, Ummay and Zhu, Haotian and Xu, Bowen and d'Amorim, Marcelo},
title = {A Case Study of LLM for Automated Vulnerability Repair: Assessing Impact of Reasoning and Patch Validation Feedback},
year = {2024},
isbn = {9798400706851},
publisher = {Association for Computing Machinery},
address = {New York, NY, USA},
url = {https://doi.org/10.1145/3664646.3664770},
doi = {10.1145/3664646.3664770},
booktitle = {Proceedings of the 1st ACM International Conference on AI-Powered Software},
pages = {103–111},
numpages = {9},
location = {Porto de Galinhas, Brazil},
series = {AIware 2024}
}

@misc{nong2025appatchautomatedadaptiveprompting,
      title={APPATCH: Automated Adaptive Prompting Large Language Models for Real-World Software Vulnerability Patching}, 
      author={Yu Nong and Haoran Yang and Long Cheng and Hongxin Hu and Haipeng Cai},
      year={2025},
      eprint={2408.13597},
      archivePrefix={arXiv},
      primaryClass={cs.CR},
      url={https://arxiv.org/abs/2408.13597}, 
}

@misc{kasan,
  title        = "{Kernel Address Sanitizer (KASAN)}",
  author       = "{The Linux Kernel development community}",
  howpublished = "\url{https://docs.kernel.org/dev-tools/kasan.html}",
  note         = "Linux Kernel documentation (version 6.17.0-rc4)",
  year         = {2025},
  key          = {kasan}
}

\section{Appendix}

\begin{table}[!h]
\caption{Patch Correctness}
\label{tab:patch-correctness}
\centering
\begin{tabular}{lllll}
\toprule
Setup                    & LLM      &Plausible   & Helpful & Wrong \\ 
\midrule
SimpleAgent              & GPT-4o   & 8         & 5       & 13        \\
Function-Exploration     & GPT-4o   & 2         & 2       & 1        \\
\bottomrule
\end{tabular}
\end{table}

\subsection{Patch correctness}
We manually verify the plausible correctness or helpfulness 31 random patches produced by our APR using GPT-4o, as shown in Table \ref{tab:patch-correctness}. As we performed manual verification, we could not determine if a patch is fully correct. We consider a patch plausibly correct if it follows the same semantics as the ground truth patch and prevents a crash, helpful if it does not properly address the root cause but targets the correct functions and prevents a crash, and wrong if it only prevents a crash but shares little to no similarity. We find that of the 31, 10 are plausibly correct, 7 are helpful, and 14 are wrong. This indicates that it is insufficient to simply rely on observing the absence of crashes to verify the correctness of patches. Interestingly, this result is consistent with what CrashFixer reported. Our rates of plausibly correct, helpful and wrong patches are 32.23\%, 22.58\%, and 45.16\%, respectively, whereas the rates for CrashFixer are 32.91\%, 15.18\%, and 51.89\%, respectively. This small study further suggests our simpler design achieved comparable performance to the much more complex solution.

\begin{table}[!h]
\caption{Compute}
\label{tab:compute-time}
\centering
\begin{tabular}{lll}
\toprule
Setup                    & LLM             & Clock Hours \\
\midrule
kgym-bm25                & GPT-4-turbo     & 11.89 \\
kgym-oracle              & GPT-4-turbo     & 13.55 \\
kgym-bm25                & GPT-4o          & 13.71 \\
kgym-oracle              & GPT-4o          & 16.14 \\
kgym-oracle+functionwise & GPT-4o          & 26.07 \\
SimpleAgent-nobic        & GPT-4o          & 45.59 \\
SimpleAgent              & GPT-4o          & 46.47 \\
SimpleAgent+Feedback     & GPT-4o          & 113.79 \\
Function-Exploration     & GPT-4o          & 45.99 \\
SimpleAgent              & Claude Opus 4.1 & 154.60 \\
SimpleAgent              & GPT-5 Thinking  & 121.24 \\
\bottomrule
\end{tabular}
\end{table}
\subsection{Compute used for experiments}
\label{appendix:compute-subsection}
We use two machines for all tests. They are identical 56 core @ 2.3GHz, 160GB RAM, 1TB SSD. We run tests sequentially, such that a build uses all 56 cores, then 26 reproducer VMs use 52 cores and 52GB of RAM (2 cores, 2GB RAM each). The APR is very IO bound (to LLM APIs) and can be run on nearly anything. When reproducing kGym, it took 4 hours using a RTX 3060 and 400GB of space to generate BM25 indices. Table ~\ref{tab:compute-time} shows compute times. Lower testing time for kGym tests can be attributed build failures ending the test early. Long test times for GPT-5 Thinking and Claude Opus 4.1 are likely due to their APIs being overloaded and forcing request retries as they had recently released, unfortunately we do not have a way of cutting that time out. They also take time to think and respond slower than GPT-4o. Preliminary testing and testing during development was also done on the same machines. We did not record time.

\subsection{Cost of KGym and GCP}
KGym requires at least three GCP instances (scheduler, builder, reproducer), in varying shapes (2x c2-standard-16, 1x c2-standard-30) at a minimum hourly cost of \$3.23 \cite{gcp-compute-pricing}. Running the minimum amount of GCP instances allows only one build job and one reproducer job to be run simultaneously, with a biweekly cost of at least \$1087.47. This cost is unsustainable for many researchers (such as ourselves) and for intensive testing that may last multiple weeks, the money is much better spent on hardware.

\begin{figure}[h]
  \centering
  \includegraphics[width=.8\textwidth]{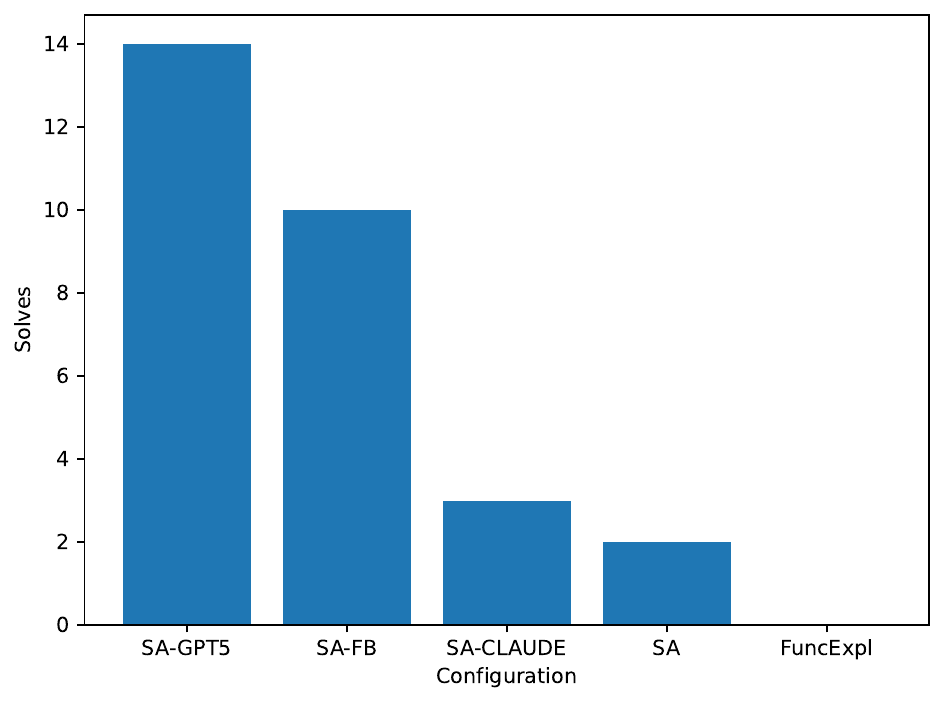}
  \caption{Unique solves per APR configuration}
  \label{fig:unique-solves}
\end{figure}

\subsection{More evaluation}
\emph{Unique Solves:} Unique solves is an interesting metric that may be helpful to show versatility. In Figure \ref{fig:unique-solves} we see the most unique solves is achieved by SimpleAgent using GPT-5, which demonstrates the unique repair capability of the model not captured by other setups using other models. SimpleAgent using GPT-4o with feedback-driven retries also proves to be capable, solving 10 bugs neither GPT-5 nor Claude Opus 4.1 solved. SimpleAgent with Claude Opus 4.1 solves only 3 unique bugs, similar to our SimpleAgent using GPT-4o, although Claude performed much better overall. Our Function-Exploration Agent collected no unique solves, although this is expected due to its low pass rate,  SimpleAgents specialization, and GPT-5's performance.

\emph{Compilation Failures:} In Table \ref{tab:build-job} compilation failures remain consistent for our agents using GPT-4o, but we see a sharp drop when using SOTA LLMs. Even Claude Opus 4.1 substantially reduces compilation failures to match GPT-5 despite not meeting the same pass rate. The reduction in compilation errors indicates both LLMs have improved capabilities to maintain internal syntactic/semantic invariants when compared to GPT-4o, even if they do not match in other aspects such as reasoning. This suggests compilation failures can be used as a proxy metric for model reliability, or at least code generation consistency.

\begin{table}[]
\caption{Reproducer Job Output}
\label{tab:reproducer-job}
\centering
\begin{tabular}{lllllll}
\toprule
Setup                    & LLM             & Pass & Trigger & Racey & Boot Fail & Other \\
\midrule
kgym-bm25                & GPT-4-turbo     & 0    & 41       & 22    & 0         & 0     \\
kgym-oracle              & GPT-4-turbo     & 2    & 36       & 18    & 1         & 0     \\
kgym-bm25                & GPT-4o          & 2    & 58       & 35    & 0         & 0     \\
kgym-oracle              & GPT-4o          & 4    & 44       & 27    & 0         & 1     \\
kgym-oracle+functionwise & GPT-4o          & 15   & 61       & 32    & 1         & 1     \\
SimpleAgent-nobic        & GPT-4o          & 25   & 85       & 57    & 5         & 0     \\
SimpleAgent              & GPT-4o          & 31   & 77       & 60    & 4         & 0     \\
SimpleAgent+Feedback     & GPT-4o          & 54   & 78       & 64    & 9         & 0     \\
Function-Exploration     & GPT-4o          & 22   & 87       & 65    & 2         & 0     \\
SimpleAgent              & Claude Opus 4.1 & 46   & 73       & 64    & 1         & 0     \\
SimpleAgent              & GPT-5 Thinking  & 62   & 60       & 60    & 0         & 0    \\
\bottomrule
\end{tabular}
\end{table}

\begin{table}[]
\caption{Build Job Output}
\label{tab:build-job}
\centering
\begin{tabular}{llll}
\toprule
Setup                    & LLM             & Compilation Fails & Bad Patch \\ 
\midrule
kgym-bm25                & GPT-4-turbo     & 7                 & 92        \\
kgym-oracle              & GPT-4-turbo     & 4                 & 63        \\
kgym-bm25                & GPT-4o          & 2                 & 78        \\
kgym-oracle              & GPT-4o          & 2                 & 55        \\
kgym-oracle+functionwise & GPT-4o          & 16                & 13        \\
SimpleAgent-nobic        & GPT-4o          & 26                & 2         \\
SimpleAgent              & GPT-4o          & 24                & 7         \\
SimpleAgent+Feedback     & GPT-4o          & 41                & 7         \\
Function-Exploration     & GPT-4o          & 24                & 8         \\
SimpleAgent              & Claude Opus 4.1 & 15                & 8         \\
SimpleAgent              & GPT-5 Thinking  & 15                & 6        \\
\bottomrule
\end{tabular}
\end{table}

\subsection{Background}
\label{sec:background}

\subsubsection{Syzkaller}
Syzkaller is an open-source coverage-guided kernel fuzzer developed by Google. It is designed to automatically discover security vulnerabilities, crashes, and unexpected behaviors in operating system kernels, with a primary focus on the Linux kernel, but it has also been adapted to other kernels like FreeBSD, NetBSD, Fuchsia, Darwin, and Windows. When a bug is found, Syzkaller is capable of outputting a reproducer program as a syz program and converting that syz program to a C program. These reproducers ideally can trigger the bug, although the reliability of the reproducer tends to vary, especially in the case of race conditions. Syzkaller has led to the discovery and reporting of thousands of Linux kernel bugs on a platform called Syzbot.

\subsubsection{Syzbot}
Syzbot is an automated bug reporting system built on top of Syzkaller and is also built by Google. Syzbot takes care of automatically triaging, reporting, and tracking bugs. It was created to reduce the manual effort needed in handling the large volume of crashes Syzkaller uncovers. Each bug entry in Syzbot has a unique ID, life cycle status (open, fixed, invalid), reproducers produced (if any), config for building, git commit, and sanitizer reports for each crash that occurs. Additionally, when the bug is fixed, the bug entry also contains the patch commit and occasionally the blamed bug inducing commit. Syzbot contains over 6500 fixed bugs and over 1500 open bugs for just the Linux kernel. This makes Syzbot an ideal source of bugs to create a benchmark.

\subsubsection{kGym} kGym is similar RGym. The project introduces a gym, a dataset, and a basic APR. kGym itself is a kernel gym for automatically testing patches. It can orchestrate compiling kernels, applying patches, and running reproducers. kGym is highly dependent on GCP (Google Cloud Platform) as tests are run on GCP virtual machines. kGym's reliance on GCP makes it easily scalable, but impossible to run locally where compute is magnitudes cheaper. kGym's baseline APR operates in two modes. Assisted (or oracle) which uses the files from the accepted patch and unassisted which uses BM25 to retrieve files relevant to the bug. Unassisted and assisted modes achieve 0.72\% and 5.38\% pass rates on their benchmark dataset respectively.

\end{document}